\documentclass[12pt]{article}
\usepackage{mathtools}
\usepackage{amsmath, amsthm, amssymb, multirow, amsfonts, amsopn, bm, bbm}
\usepackage{subcaption}
\usepackage{relsize} 
\usepackage{hyperref,tikz,epsfig}
\usepackage[edges]{forest}
\usepackage{tikz}
\usetikzlibrary{trees}
\usepackage{mwe}
\usepackage{breqn} 
\usepackage{forest}
\usepackage{graphicx} 
\usepackage[margin=2.5cm]{geometry}
\usepackage{setspace}
\usepackage{url}
\usepackage{authblk}
\usepackage{graphicx}

\title{\large Long-Term Mortality Following STN-DBS in Parkinson’s Disease: A Survival Analysis}

\author[1]{\small Malinda Iluppangama}
\author[2*]{\small Dilmi Abeywardana}
\affil[1]{\small Department of Mathematics and Statistics, Loyola University Maryland, Baltimore, MD, USA.}
\affil[2]{\small Department of Mathematics and Statistics, University of South Florida, Tampa, FL, USA.}
\affil[*]{\small Corresponding author: Dilmi Abeywardana (dabeywardana@usf.edu)}

\date{}

\begin{document}

\maketitle
\vspace{-1.5cm}
\begin{abstract}

\textbf{Background:} Deep brain stimulation of the subthalamic nucleus (STN-DBS) is an established PD treatment that improves motor symptoms and quality of life, but long-term survival and mortality risk remain poorly characterized. This study evaluated long-term survival and mortality-associated factors after STN-DBS.

\textbf{Methods:} A retrospective survival analysis of 214 PD patients who underwent STN-DBS used Kaplan-Meier methods to estimate survival probabilities and median survival, with log-rank tests comparing survival by gender and age at implantation. Cox proportional hazards (Cox-PH) models identified mortality-associated factors in both univariate and multivariate analyses, and were validated using accelerated failure time (AFT) models.

\textbf{Results:} Mortality rose progressively, with 8.0\%, 42.5\%, and 56.0\% of patients dying within 5, 10, and 15 years, respectively; median survival was approximately 134 months. Survival did not differ by gender (p = 0.63) but was significantly lower for patients aged $\geq$60 years at implantation (p $<$ 0.0001). At least one post-implantation revision reduced mortality risk (HR = 0.296, p $<$ 0.001), while implantation at age $\geq$60 years increased it (HR = 4.689, p $<$ 0.001); both remained significant after adjustment and were confirmed by AFT models. Patient satisfaction was high, with 97\% of 52 surveyed patients reporting satisfaction and 96\% would recommend STN-DBS.

\textbf{Conclusion:} STN-DBS is linked to long-term survival and high patient satisfaction in individuals with PD. Older age at implantation increased mortality risk, whereas revision surgery and mortality likely reflect confounding by indication rather than a direct causal effect. These findings support age-stratified patient selection and highlight the importance of continued long-term monitoring following STN-DBS.\newline


\end{abstract}

\section{Introduction}
\label{intro}
Parkinson's Disease (PD) is the most prevalent neurodegenerative disorder that affects the brain neuron system after Alzheimer's disease, and remains a major global health concern despite decades of research to understand its cause by medical professionals and research scientists; yet, the cause of PD remains undetected. According to the World Health Organization (WHO), more than 10 million people worldwide currently live with PD, a number projected to reach 25.2 million by 2050. The WHO further estimates that neurodegenerative diseases, including PD and Alzheimer's disease, will surpass cancer to become the second leading cause of death worldwide by 2040,~\cite{PD_projection},~\cite{PD_Future_deaths}. In the United States alone, more than 90,000 new cases are diagnosed each year, with an estimated 1 to 1.5 million people living with the disease, figures that likely underestimate the true burden given the large number of undiagnosed cases worldwide,~\cite{PD_challange}.

PD is broadly classified into two main types: idiopathic and genetic. Genetic PD is a comparatively rare form caused by inherited gene mutations and often has an earlier age of onset, while idiopathic PD, the far more common form, has no identifiable cause. Patients diagnosed before the age of 50 are classified as having young-onset PD. The main symptoms of PD are motor in nature, including tremor, rigidity, bradykinesia, and postural instability, although the disease also carries a substantial non-motor burden. Symptoms of PD typically begin slowly and progress gradually over time, and patients in the later stages often experience severe impairment in walking and speech. Beyond its motor symptoms, PD is associated with a range of non-motor complications, including cognitive decline and, in a substantial proportion of patients, the eventual development of dementia, further compounding the impact of the disease on patients' independence and quality of life~\cite{PD_cause},~\cite{Rmayya_dbs}.
Research scientists have been working over the past few decades and have not found a cure for this burdensome disease. During the past several decades, researchers have developed a range of therapies to slow the progression of PD and alleviate its symptoms. Levodopa and dopamine agonists remain the most commonly prescribed medications, helping to manage motor symptoms in the early stage of the disease~\cite{Levodopa_PD_medication}. However, as PD advances, these drug therapies alone often become insufficient to control symptoms, prompting clinicians to consider surgical intervention. Deep Brain Stimulation (DBS) has become one of the most effective treatments for advanced PD, delivering targeted electrical stimulation to specific brain regions to improve motor function and overall quality of life. Because movement disorders represent one of the primary symptom categories of PD, DBS has become widely adopted in its treatment, showing particularly strong outcomes for bradykinesia, tremor, and other motor complications. Beyond symptom control, DBS has also been shown to reduce patient dependence on medication, offering an additional clinical benefit in conjunction with its direct effects on motor function~\cite{parkinsonsfoundation},~\cite{PD_dbs_imporve},~\cite{Deuschl_et_al_DBS_improvement}.
 
Several studies in the literature confirm that DBS therapy improves survival times for PD patients compared to other treatments,~\cite{Ngoga_DBS_survival}. Multiple studies examining long-term survival outcomes of PD patients who underwent DBS consistently report a significant improvement in motor symptoms and quality of daily life. DBS has also been shown to reduce medication needs, with a daily equivalent dose of levodopa that decreased by more than 55\% at 5-year follow-up,~\cite{Fasano_8_year_dbs_followup},~\cite{5_year_folloup_DBS}. The survival rates reported in the literature are similarly encouraging: several studies report survival rates above 98\% at 5 years and very high survival even at 3 years after DBS,~\cite{Kim_mortality_DBS},~\cite{Ryu_DBS_mortality},~\cite{Kim_A_DBS_mortality}, while one study estimated long-term (10-year) survival probability at 51\%,~\cite{Rmayya_dbs}.

Despite these promising findings, most of this literature shares a common statistical limitation. The majority of studies rely only on the Kaplan-Meier method, the most widely used approach for survival analysis, but one that is neither especially powerful nor as robust as parametric or semi-parametric methodologies. Many studies also use the semi-parametric Cox proportional hazards model without stating whether its underlying assumptions actually hold, meaning their conclusions about patient survival may be unreliable when those assumptions are violated.

In this study, we build on our previous work applying machine learning to the detection of PD~\cite{PDIluppangama2025} and statistical modeling of Parkinson's disease~\cite{Encyclopedia_PD},~\cite{iluppangama2025_survival}, extending our methodological focus to the long-term survival outcomes of patients with PD who underwent DBS. Rather than relying on a single statistical approach, we utilize a comprehensive framework combining nonparametric, semi-parametric, and robust parametric survival analysis to characterize these outcomes more rigorously than prior single-method studies have allowed. Specifically, our objective is to (1) characterize and compare the underlying survival time distributions of male and female patients using Kaplan-Meier estimation, (2) identify clinically significant prognostic factors, including age at implant, gender of patients, and number of revision surgeries, using Cox Proportional Hazards regression and Accelerated Failure Time models. The remainder of this paper is organized as follows: Section 2 presents a data description and descriptive analysis of the data; Section 3 introduces the statistical methodologies employed; Section 4 presents the results of the analysis; and Section 5 concludes the study, highlighting its contributions and clinical implications.

\section{Data Description and Descriptive Analysis}
\label{sec:1}
The data consist of information on the first 400 individuals who underwent DBS treatment for PD between 1999 and 2007 at the Pennsylvania Hospital, University of Pennsylvania,~\cite{Rmayya_dbs}. The data comprise 320 individuals who underwent DBS for PD, and the demographics of the cohort are as follows: the mean age of the DBS implant was approximately 61($\pm$ 9.44) years, 70\% of the patients were male, and during the follow-up period, approximately 54\% of the patients died at an approximate mean age of 73 ($\pm$8.5) years. Furthermore, approximately 78\% of patients underwent subthalamic nucleus (STN-DBS) and approximately 73\% of patients underwent bilateral implantation. Furthermore, these data are consistent with the results of the follow-up telephone survey of 52 patients in this cohort, and 94\% of them were happy with their STN-DBS and 96\% of them would recommend STN-DBS.

For survival analysis, we have selected individuals who performed bilateral STN-DBS and ended with n=214 patients; all of them were included in the univariate survival analysis using non-parametric estimation. mean age of deceased is approximately 72($\pm$8.4) years and on average approximately patients in this subset alive 96($\pm$39) months after DBS implantation.

Finally, for multivariate survival analysis using semi-parametric and fully parametric methods, we have included age at implant, gender, and number of revisions,~\cite{PD_age},~\cite{Rmayya_dbs}. For this subset, we had to remove several observations from the analysis due to missing values. So, for the multivariate analysis, we only utilized 189 patients. Furthermore, Age at Implant as categorical variable with three levels (less than 50, between 50 and 59 and greater than 60), approximately 54\% of the patients belongs to 50 to 59 age category, Gender of the patient (Male or Female), approximately 72\% of the patients were male and number of revisions each patient has completed as a categorical variable with two level (0 revision or one or more revisions) and 79\% of the patients were undergo at least one revision.

\section{Methodology}
\label{sec:2}

In this study, we utilize the well-known non-parametric  Kaplan-Meier (KM) approach, a robust parametric accelerate faliure time (AFT) approach, and the widely used semi-parametric Cox-PH method to analyze the long-term survival outcomes of individuals who underwent STN-DBS for PD. In the following section, the proposed models and their usefulness are briefly discussed.

\subsection{Kaplan-Meier Estimation(KM)}
\label{sec:2:KM_method}

Kaplan-Meier(KM) estimation of the survival time is a fully non-parametric approach for analyzing the time of occurrence of the event of interest (time until death), and it was first introduced in 1958,~\cite{KM}. KM is a well-known approach in the field of survival analysis and reliability analysis. However, KM is not as powerful or robust as parametric or Bayesian survival analysis since it does not make distributional assumptions about the underlying survival time. KM involves estimating the survival probabilities of patients at a certain point in time as follows,

\begin{equation*}
    S_t = \frac{\text{Total patients living at time } t_0 - \text{Total patients who died by time } t}{\text{Total patients living at time } t_0}
\end{equation*}

Suppose that there are $n$ individual patients at the start of time ($t_0$) and the respective observed event times (died at time $t_i$) are $t_1 < t_2 < t_3 <...<t_k$ with $d_i$ units failing at time $t_i$. Then the KM estimator of the survival function $S_{KM}(t)$ is given by the following equation,

\begin{equation}
    S_{KM}(t) = \prod_{i:t_i<t} \frac{(n_i - d_i)}{n_i} = \prod_{i:t_i<t}(1-\frac{d_i}{n_i})
    \label{Eq_KM},
\end{equation}

where $t>0$,
$n_i$ is the number of patients at risk at time $t_i$,
$d_i$ is the number of patients who died at time $t_i$. 

We proceed and calculate survival probabilities using KM estimators given by Equation~\ref{Eq_KM} under the following assumptions,~\cite{KM_Basics},~\cite{KM_Explain},

\begin{itemize}
    \item Patients who have been censored have the same survival prospects as those who continue follow-up.
    \item Survival probabilities are the same for individuals who were recruited early and late in the study.
    \item Events happened at the specified time.
\end{itemize}

\subsection{Multivariate Cox-Proportional Hazards Model(Cox-PH)}
\label{sec:2 COXPH}

In the context of survival/reliability analysis, the hazard function plays a major role. It explains the instantaneous risk of an event at time $t$. That is, the probability that the desired event will occur any time after time $t$ is the hazard function under the assumption that the individual observation has not experienced the desired event until time $t$.\\

Let $T$ be a random variable that represents the survival time of a given patient. Then the probability that an event occurs prior to or at time $t$ is defined by the cumulative distribution function (CDF) and is given below,

\begin{equation}
    F(t) = P(T \leq t) = \int_{0}^{t} f(t)\,dt\quad, \quad\quad t>0,
    \label{Cumulative}
\end{equation}
 where the probability density function can be derived as follows,
\begin{equation}
    f(t) = \frac{dF(t)}{dt}.
    \label{probability_function}
\end{equation}
The probability of a given patient surviving after time $t$ (Survival function) can be defined by the following Equation~\ref{survival_function}, 

\begin{equation}
    S(t) = 1 - F(t) = P(T > t) = \int_{t}^{\infty} f(t)\,dt\quad, \quad\quad t>0.
    \label{survival_function}
\end{equation}

Using Equation~\ref{probability_function} and Equation~\ref{survival_function}, the hazard function for an individual patient at time $t$ can be derived as follows,

\begin{equation}
    h(t) = \lim_{\delta t\to\ 0} \frac{P(t < T \leq t +\delta t | T > t)}{\delta t} = \frac{f(t)}{S(t)}.
    \label{hazard_fun_def}
\end{equation}

To estimate the hazard function, the Cox Proportional Hazards Model(Cox-PH) was introduced by D. R Cox in 1972,~\cite{COX}, which specifies the hazard function directly rather than deriving it as the ratio of the probability density function and the survival function. The analytical form of the Cox-PH model is given by Equation~\ref{coxph_model},

\begin{equation}
\label{coxph_model}
    h_i (t) = h_0 (t) exp\left(\sum_{j=1}^{p} \beta_j X_{j} \right ),
\end{equation}

where $h_0(t)$ is the baseline hazard function, which depends only on the survival time. $\beta_j$ is the coefficient parameter of $j^{th}$ covariate. In the Cox-PH model, the baseline hazard function $h_0 (t)$ is unspecified and is treated as a non-parametric function of time $t$. However, the Cox-PH model assumes the time independence of the covariates, linearity in the covariates, additivity, and proportional hazards. Thus, the Cox-PH model is considered as a semi-parametric survival model,~\cite{Abeywardana2025_Cox_PH}. To estimate the parameters of the Cox-PH model, we cannot use the full likelihood approach since the baseline hazard function is not specified. Thus, the partial likelihood approach proposed by Cox in 1979 is utilized to estimate the model coefficients of the Cox-PH model,~\cite{Cox_theory}. 

In many situations, researchers are interested in the factor exp(coef) of the model rather than estimating the full $h_i(t)$, which determines how an individual patient varies relative to the baseline hazard function. Equation~\ref{coxph_model} can be rearranged to find the hazard ratio for a given individual patient as given below,

\begin{equation}
   HR = \frac{h_i(t)}{h_0(t)}  = exp\left( \sum_{j=1}^{p} \beta_j X_{ij}\right).
    \label{hazard_ratio}
\end{equation}

Suppose we have two types of treatment groups: the treatment group(T) and the placebo group(P). Then we can extract the following important and useful information about the two treatment groups using the HR,

\begin{itemize}
    \item HR $> 1 (\beta_i>0$). The treatment group has a higher hazard than the placebo group. (i.e., the placebo group is favored.)
    \item HR $\approx1$. The treatment group and placebo group have no significant difference.
    \item HR $< 1 (\beta_i<0$). The treatment group has a lower hazard than the placebo group. (i.e., the treatment group is favored.)
\end{itemize}

\subsection{Accelerated Failure Time Model(AFT)}
\label{sec:2 AFT}

The theory of the Accelerated Failure Time (AFT) model has been an active area of research for the last few decades in the domain of time-to-event data. The AFT model is a parametric model that was introduced by Cox (1972). The AFT model provides a fully parametric framework that describes the relationship between survival time(T) and explanatory variables. In this model, the logarithms of the survival times are considered as a response variable and include an error term, which is assumed to follow a specific probability distribution. The relationship between survival time and the explanatory variable is a linear relationship between the logarithm of survival time and explanatory variables, as given by Equation~\ref{aft_model}.

\begin{equation}
    \log(T_i) = \beta_0 + \sum_{j=1}^{p}\beta_j X_{ij} + \sigma \epsilon_i,
    \label{aft_model}
\end{equation}

where $\beta_0$ is the intercept, $\beta_j$ is the coefficient parameter for the $j^{th}$ covariate, $\sigma$ is a scale parameter and $\epsilon_i$ is a random error term whose distribution is assumed to follow a specific distribution such as exponential, Weibull, log-normal, log-logistic, etc,~\cite{collett2015}.

Equivalently, the AFT model can be written directly in terms of the survival function. If $S_0(t)$ denotes the baseline survival function, the survival function for an individual with covariate vector $X_i$ is given by Equation~\ref{aft_survival},
\begin{equation}
    S_i(t) = S_0\!\left(\frac{t}{\exp(\sum_{j=1}^{p}\beta_j X_{ij})}\right),
    \label{aft_survival}
\end{equation}
which shows explicitly that covariates act to rescale, or ``accelerate'' the time axis relative to the baseline survival function~\cite{lawless2003}.

Unlike the Cox-PH model, where the baseline hazard function $h_0(t)$ is not specified, the AFT model requires a distributional assumption for $T$, and model parameters $\beta_j$ and $\sigma$ are estimated with full maximum likelihood estimation,~\cite{kalbfleisch2002}. This distributional assumption gives the AFT model greater statistical efficiency when correctly specified, but makes it more sensitive to model mis-specification than the semi-parametric Cox-PH model,~\cite{collett2015},~\cite{AFT_Method}. From the coefficients of the AFT model, it is possible to define the "Time Ratio" (TR), calculated as the exponential of the coefficient, as shown in Equation~\ref{time_ratio}. 

\begin{equation}
    TR_j = \exp(\beta_j).
    \label{time_ratio}
\end{equation}

The TR is easier to interpret than the raw coefficient, and its formal interpretation is as follows. Suppose that we have a treatment group and a placebo group. Then the following interpretation of the TR applies:
\begin{itemize}
    \item TR $> 1$ $(\beta_j > 0)$. The effect of treatment is decelerated (The treatment group has a longer expected survival time than the placebo group.
    \item TR $\approx 1$. The treatment group and the placebo group have no significant difference in survival time.
    \item TR $< 1$ $(\beta_j < 0)$. The effect of treatment is accelerated (The treatment group has a shorter expected survival time than the placebo group.
\end{itemize}

\section{Results-Survival Analysis}
\label{sec:3}

In the following subsection, we summarize the results of survival outcomes of KM, Cox-PH, and AFT for PD patients who underwent STN-DBS for PD.

\subsubsection{Kaplan-Meier(KM)}
\label{sec:3 KM}

\begin{figure}[h]
    \centering
    \includegraphics[width=5in]{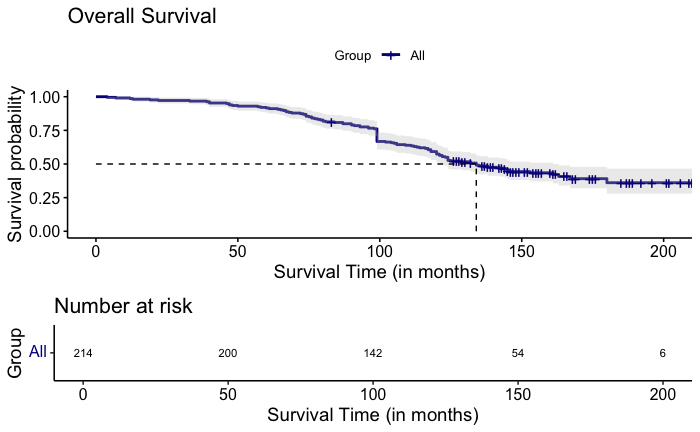}
    \caption{Overall Survival.}
    \label{fig:KM_all}
\end{figure}

First, we analyzed overall survival after DBS in all patients without stratification using Kaplan-Meier(KM) estimation as described in~\ref{sec:2:KM_method}. The resulting survival curve and the corresponding risk table are shown in Figure~\ref{fig:KM_all}. The KM-estimated overall median survival was 134 months (95\% CI:(122,163)) for the entire STN-DBS surgery for PD cohort. This initial analysis included all patients as a single group, without stratifying by baseline patient characteristics. 

Next, we investigated the association between several baseline characteristics and survival time in this cohort. These baseline characteristics included age group at STN-DBS surgery, sex, and number of revision surgeries after the initial procedure (we have categorized the number of revisions into two groups as at least one surgery vs no surgery after the initial surgery). KM survival estimation was used to characterize survival across these baseline characteristics. To formally compare survival distributions between groups, we performed the log-rank
test,~\cite{log_rank},~\cite{Wilcox_rank_sum}. Statistical significance was set at p $< $0.05 (two-sided).

\begin{figure}[ht]
    \centering
    \includegraphics[width=5in]{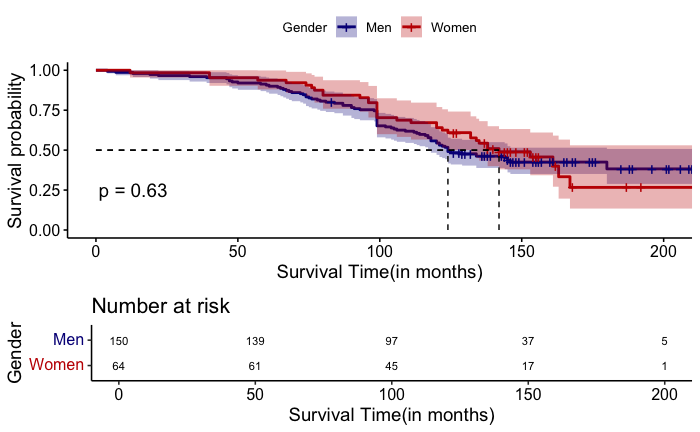}
    \caption{Overall Survival Stratified by Gender.}
    \label{fig:KM_Gender}
\end{figure}

We first examined the effect of gender on patient survival after STN-DBS surgery. The survival curves of KM stratified by gender are presented in Figure~\ref{fig:KM_Gender}. To determine whether survival differs between male and female patients, the following hypotheses were tested:

\begin{center} $H_0$: The survival distributions are the same for male and female patients.\\ vs

$H_1$: The survival distributions differ between male and female patients. \end{center}

Figure~\ref{fig:KM_Gender} provides a graphical comparison of the estimated survival functions for males and females, while the accompanying statistical tests quantify whether any observed differences are statistically significant. The outcome of the hypothesis testing has a p-value of 0.63.

\begin{figure}[ht]
    \centering
    \includegraphics[width=5in]{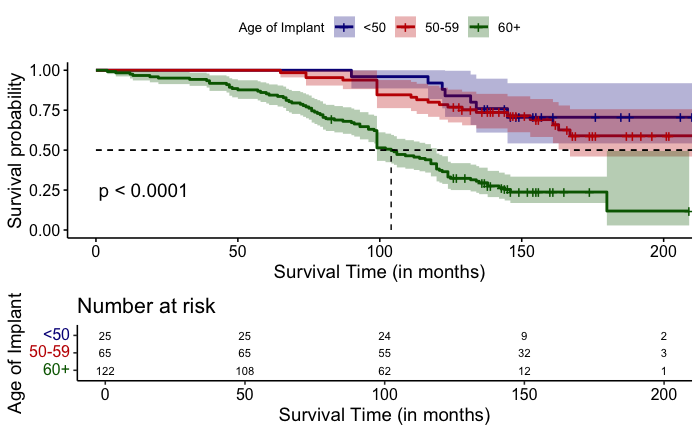}
    \caption{Overall Survival Stratified by Age of Implant.}
    \label{fig:KM_age_group}
\end{figure}

Subsequently, we evaluated the impact of the age group at STN-DBS implantation. Patients were categorized into three age groups and KM curves were estimated for each group. The resulting survival curves are shown in Figure~\ref{fig:KM_age_group}. To assess whether survival experiences differ across age categories, the following hypotheses were considered:

\begin{center} $H_0$: Survival distributions are identical across all three age groups.\\ vs

$H_1$: At least one age group has a survival distribution that differs from the others. \end{center}

The KM curves provide a visual assessment of differences in survival probabilities over time, while the log-rank and Wilcoxon tests formally evaluate whether these differences are statistically significant. Together, these analyzes help identify whether age at implantation is associated with variation in long-term survival outcomes following DBS. The resulting p-value is $<$ 0.0001.

\subsubsection{Cox-PH Model}
\label{sec:3 Cox-PH results}

To evaluate the association between baseline characteristics and survival outcomes, we conducted univariate and multivariate Cox proportional hazards (Cox-PH) analysis on a cohort of 189 patients. The estimated regression coefficients, the hazard ratios, and the associated statistical measures are summarized in Table~\ref{Coefficient_ table for Coxph Model}.

\begin{table}[ht]
\caption{Coefficients and hazard ratios of univariate and multivariate Cox proportional-hazards models in female patients who underwent STN-DBS.}
\begin{center}
\begin{tabular}{ccccc}
    \hline   
      Factor & Coefficient  & Hazard Ratio (exp(coef)) & P-value \\
     \hline
     & \textbf{Univariate Cox-PH} & & & \\
     \hline
     Gender(Female) & 0.005 & 1.005(0.655, 1.54) & 0.984\\
     
     Age at Implant($50-59$) & 0.189  & 1.209(0.511, 2.862)  &0.666\\

     Age at Implant($\ge60$) & 1.545  & 4.689(2.150, 10.226) & $<0.001$\\
     
     Number of Revisions($\ge1$) &  -1.224 & 0.294(0.194, 0.445) & $<0.001$\\

     \hline
    & \textbf{Multivariate Cox-PH} & & &  \\
     \hline
     Gender(Female) & -0.094 & 0.911(0.592, 1.401)& 0.670 \\
     
     Age at Implant($50-59$) & 0.272 & 1.313(0.552, 3.122) & 0.537\\
     Age at Implant($\ge60$) & 1.484 & 4.411(2.019, 9.641) & $<0.001$ \\
     Number of Revisions($\ge1$) & -1.039 & 0.354(0.233, 0.537) & $<0.001$\\
     \hline
\end{tabular}
\end{center}

\label{Coefficient_ table for Coxph Model}
\end{table}

\begin{figure}[ht]
    \centering
    \begin{subfigure}{0.45\textwidth}
        \centering
        \includegraphics[width=\textwidth]{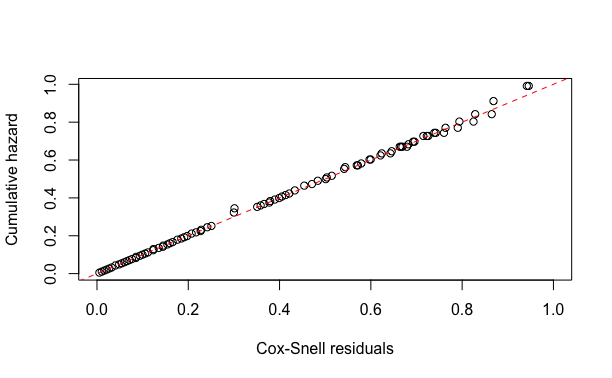}
        \caption{Cox-Snell Plot for Gender}
        \label{fig:1a}
    \end{subfigure}
    \hfill
    \begin{subfigure}{0.45\textwidth}
        \centering
        \includegraphics[width=\textwidth]{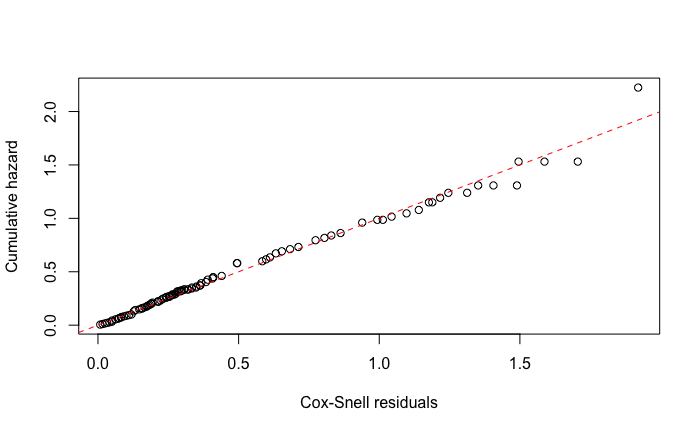}
        \caption{Cox-Snell Plot for Age at Implant}
        \label{fig:1b}
    \end{subfigure}
    
    
    \begin{subfigure}{0.45\textwidth}
        \centering
        \includegraphics[width=\textwidth]{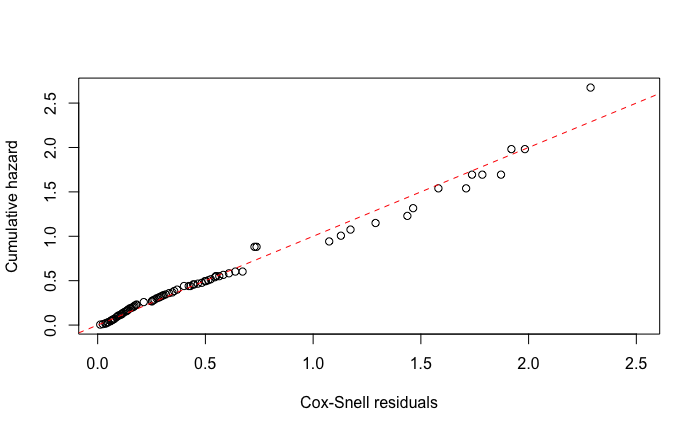}
        \caption{Cox-Snell Plot for N Revisions}
        \label{fig:1c}
    \end{subfigure}
    \hfill
    \begin{subfigure}{0.45\textwidth}
        \centering
        \includegraphics[width=\textwidth]{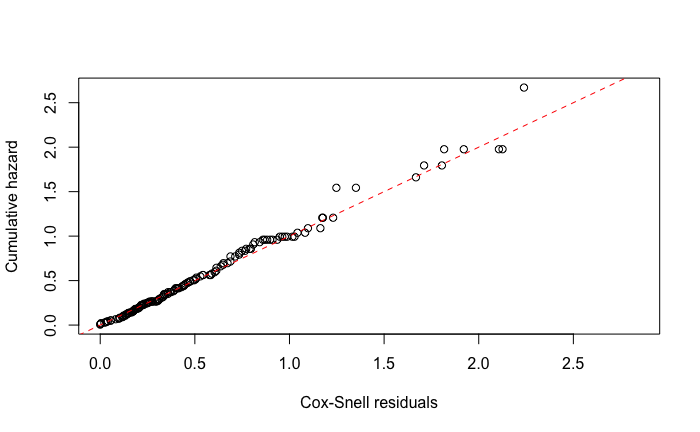}
        \caption{Cox-Snell Plot for Multivariate model}
        \label{fig:1d}
    \end{subfigure}
    
    \caption{Cox-Snell plots to check overall model adequacy.}
    \label{fig:1}
\end{figure}

Furthermore, we assessed the adequacy of the Cox proportional hazards (Cox-PH) model by examining its underlying assumptions, including the proportional hazards assumption. The results indicated that none of the covariates violated the proportional hazards assumption, suggesting that the model is appropriate for the data. The overall goodness-of-fit of the model was further evaluated using Cox-Snell residual analysis, which demonstrated a satisfactory fit as shown in Figure~\ref{fig:1}. Furthermore, to assess the predictive performance and robustness of the multivariate Cox-PH model, we performed a 5-fold cross-validation and calculated the concordance index(C-index). The model achieved a C-index of 0.73 ($\pm$ 0.018) which attests to the robustness of the multivariate Cox-PH model.

\subsubsection{Accelerated Failure Time Model (AFT)}
\label{sec:3 AFT results}

A log-logistic accelerated failure time (AFT) model was fitted to assess the association between gender, age group at STN-DBS surgery and number of revision surgeries with survival time. The log-logistic distribution was selected over Weibull, lognormal, and exponential alternatives based on the lowest Akaike Information Criterion (AIC) (AIC = 1207.4 vs. 1208.9, 1221.6, and 1281.8, respectively). The estimated model coefficients, time ratios, and associated statistical measures are summarized in Table~\ref{Coefficient_ table for AFT Model}.

\begin{table}[ht]
\caption{Coefficients and time ratios of the accelerated failure time (AFT) model.}
\begin{center}
\begin{tabular}{ccccc}
    \hline   
      Factor & Coefficient  & Time Ratio (95\% CI) & P-value \\
     \hline
     Gender(Female) & 0.034 & 1.035 (0.856, 1.250)& 0.720\\
     
     Age at Implant(50-59) & -0.134	  & 0.875 (0.629,1.214)&  0.422\\
     Age at Implant($\ge60$) & -0.693  & 0.500(0.368,0.679)  &  $<0.001$\\
     
     Number of Revisions($\ge1$) &  0.519	 & 1.679(1.365, 2.066) &  $<0.001$\\

     \hline
\end{tabular}
\end{center}
\label{Coefficient_ table for AFT Model}
\end{table}

\begin{figure}[ht]
    \centering
    \includegraphics[width=5in]{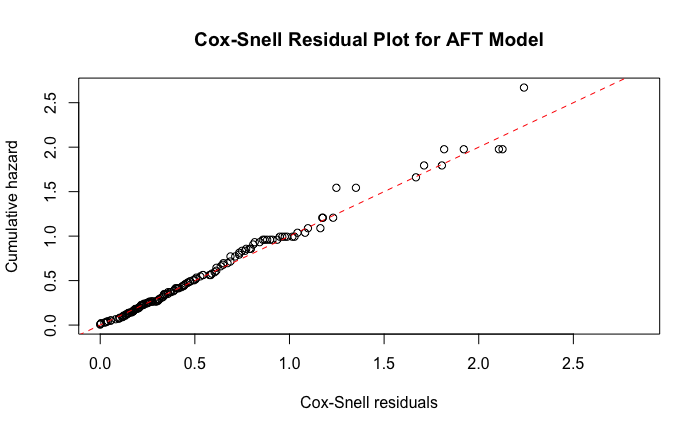}
    \caption{Cox-Snell plots to check overall model adequacy.}
    \label{fig:CS_AFT}
\end{figure}

Furthermore, Cox-Snell residual analysis confirmed an adequate model fit, with residuals closely following the expected 45° reference line shown in Figure~\ref{fig:CS_AFT}.

\section{Discussion and Conclusion}
\label{sec:4}

A total of 214 patients with PD who underwent STN-DBS were included in the survival analysis. Over the course of follow-up, mortality increased steadily. Specifically, 18 patients (8\%) died within the first 5 years after surgery, while 92 patients (42.5\%) and 121 patients (56\%) died within 10 and 15 years of follow-up, respectively. These findings highlight the long-term survival outcomes of patients treated with STN-DBS surgery and provide important information regarding mortality over extended follow-up periods. In addition to this information, it is consistent with outcomes of the follow-up telephone survey of 52 patients from this cohort, and 97\% of them were happy with their STN-DBS, and 96\% of them would recommend STN-DBS.

Next, we estimated survival outcomes for patients who underwent STN-DBS for PD using the widely adopted Kaplan-Meier (KM) estimator, based on a cohort of 214 patients. We first performed KM estimation using all patients without any stratification; the graphical visualization is shown in Figure~\ref{fig:KM_all}, and the estimated median survival time for patients who underwent STN-DBS for PD was approximately 134 months (95\% CI: 122 163 months).

An important consideration is that this unstratified KM estimation accounts only for survival time and ignores all baseline characteristics. In reality, patients' baseline characteristics differ from one patient to another, such as gender, age at implant surgery, and other factors. To avoid the bias of treating all patients as having identical baseline characteristics, we therefore performed KM estimation together with hypothesis testing on the gender and age group at implant variables.

Graphical visualizations of the KM estimates stratified by these two baseline characteristics: gender and age at implant are given in Figure~\ref{fig:KM_Gender} and Figure~\ref{fig:KM_age_group}, respectively. We additionally performed statistical hypothesis testing on these two variables to assess whether the distribution of survival time differs across the levels within each variable. For the gender variable, the estimated p-value was 0.63($>0.05$), which is not statistically significant, and we therefore do not have enough evidence to conclude that the survival time distribution of male patients who underwent STN-DBS for PD is not significantly different from that of female patients at the $5\%$ level of significance. In contrast, for the hypothesis test, the estimated p-value for the age group at the implant variable was $<0.0001$ ($<0.05$), which is statistically significant, so here at the $5\%$ level of significance we have enough evidence to conclude that the survival time distribution of patients who underwent STN-DBS differs significantly across age groups. Furthermore, as shown in Figure~\ref{fig:KM_age_group}, patients aged 60 years or older showed significantly lower survival probability compared with patients in other age groups. Furthermore, these findings are consistent with the findings we found in the literature,~\cite{PD_age}.

Non-parametric methods typically make no distributional assumptions about the underlying survival time or hazard function, and they do not incorporate covariate information into their estimations or do not apply to predictive modeling. Because of these limitations, KM estimates are less informative than semi-parametric or parametric approaches. We therefore used both a semi-parametric method (Cox-PH) and a parametric method (AFT) to analyze mortality among the patient cohort who underwent STN-DBS for PD; the results for Cox-PH and AFT models are presented in Section~\ref{sec:3 Cox-PH results} and~\ref{sec:3 AFT results}, respectively.

In the univariate Cox-PH analysis, at least one revision surgery following the initial STN-DBS implant was associated with a hazard ratio of 0.296, and this effect was statistically significant (p $<<$ 0.001). This indicates that patients who underwent one or more revisions had substantially lower mortality risk than those who did not. In contrast, the age group at implant showed that patients aged 60 years or older had a significantly higher hazard ratio (4.689) compared with younger age. This provides strong evidence that older age at implant is associated with poorer survival among patients undergoing STN-DBS for PD, relative to younger patients at implant. We additionally performed a multivariate mortality analysis using a Cox-PH model, which confirmed the findings from the univariate Cox-PH: a higher number of revision surgeries was associated with reduced mortality, and an implant age of 60 years or older was associated with increased mortality, after adjusting for other covariates. Cox-PH models are reliable only if their underlying assumptions hold; thus, we evaluated all model assumptions, including proportional hazards, and found no violations. We further assessed the overall goodness-of-fit of each model using Cox-Snell residual plots; all plots consistently indicated that the fitted models provide a reasonable fit to the observed data, and under 5-fold cross-validation, C-index = 0.73($\pm0.018$), which illustrates the high quality of the model.

Finally, according to the log logistic AFT model results, which are reported in section~\ref{sec:3 AFT results}, both the number of revision surgeries ($\ge1$) and age group at STN-DBS implant were statistically significant with baseline characteristics. For age group at implant, the estimated TR was Tr$=exp(-0.693)=0.50<1$. This result suggests that patients who received the implant at the age of 60 or older had shorter survival time compared to those who received iSTN-DBS at the age of 50 years or younger. Specifically, the expected median survival time for the 60 or older age group was approximately $50\%$ of that of the 50 years or younger. For the number of revision surgeries, the estimated TR$=exp(0.519)=1.679>1$. Since $TR>1$, patients who underwent at least one revision surgery were expected to live longer compared to those who did not undergo any revision surgery after initial STn-DBS surgery. Specifically, the expected survival time for patients with at least one revision surgery was approximately $67.9\%$ longer than that of patients without revision surgery. Furthermore, the appropriateness of the log-logistic model was evaluated using the Akaike Information Criterion (AIC) and its overall goodness of fit was assessed through Cox-Snell residual analysis.

Finally, we compared the results of the Cox proportional hazards (Cox-PH) model with those of the log-logistic accelerated failure time (AFT) model. The direction of association for both statistically significant covariates was consistent across the two modeling approaches. Specifically, implantation at age 60 years or older was associated with an increased hazard of death in the Cox-PH model (HR = 4.411) and a shorter survival time in the AFT model (TR = 0.500). Similarly, having at least one revision surgery after the initial STN-DBS procedure was associated with a reduced hazard of death in the Cox-PH model (HR = 0.354) and a longer survival time in the AFT model (TR = 1.679). These consistent findings support the robustness of the study results. The corresponding coefficient comparison is presented in Table \ref{Comparison_Cox_AFT}. Although the hazard ratios and the time ratios are not directly comparable in magnitude, the direction of association was consistent across both modeling frameworks for all significant covariates, further reinforcing the reliability of the identified risk factors.

\begin{table}[ht]
\caption{Comparison of Survival Outcomes between Cox-PH vs Log-Logistic AFT}
\begin{center}
\begin{tabular}{cccccc}
    \hline  
     & Cox-PH & Cox-PH & AFT & \\
     \hline
      Factor & Hazard Ratio  & Hazard Ratio & Time Ratio &\\
     \hline
     & \textbf{Univariate} & \textbf{Multivariate} & \textbf{Log-Logistics} &  Significant\\
     \hline
     Gender(Female) & 1.005 & 0.911 & 1.035 &\\
     
     Age at Implant($50-59$) & 1.209  & 1.313  &0.875 &\\

     Age at Implant($\ge60$) & 4.689 & 4.411 & 0.500 & **\\
     
     Number of Revisions($\ge1$) &  0.294 & 0.354 & 1.679 & ***\\
     \hline
\end{tabular}
\end{center}
\label{Comparison_Cox_AFT}
\end{table}

This study reports several robust findings, such as the number of revision surgeries and age group at implant. These findings were supported by consistent model diagnostics and robust evidence for their clinical relevance. These should be considered when assessing individual patient risk and guiding postoperative monitoring following STN-DBS surgery for PD. The robustness of these results stems in part from the use of parametric survival analysis, an approach that remains relatively uncommon in the STN-DBS survival literature. Because the underlying parametric assumptions were satisfied, the resulting models offer greater efficiency and robustness compared with non-parametric and semi-parametric methods. In addition, all relevant assumptions, including the proportional hazards assumption, were formally assessed, a step that is infrequently reported in survival analyses of STN-DBS patient cohorts. Finally, this study benefits from a large and well-characterized patient cohort who underwent STN-DBS surgery for PD.

Despite these strengths and promising results, this study has limitations in assessing mortality risk after STN-DBS surgery. In particular, the analysis relied on a limited set of clinical and demographic variables and did not incorporate additional data modalities. Future studies should prioritize the collection of data across multiple sites, data centers, and countries to capture more diverse patient populations and data types. Exploring multi-modal approaches that integrate imaging, genomic, and clinical variables may further improve mortality risk assessment and enhance clinical utility. With richer and more diverse data, survival analysis approaches such as those employed here hold strong potential for clinical translation—improving risk stratification, enabling more personalized treatment strategies, and ultimately improving outcomes for patients with Parkinson's disease.

\section{Statements and Declarations}

\subsection*{Competing Interests:}

The authors declare that they have no competing interests.

\subsection*{Acknowledgement:}

We would like to thank Dr. Ashwin Ramayya and his collaborators at the Division of Clinical Research of Neurosurgery, Pennsylvania Hospital, for granting us access to the deep brain stimulation dataset used in this study.

%
 \section*{Conflict of Interest:}
The author(s) declared no potential conflicts of interest with respect to the research, authorship, and/or publication of this article.

\bibliographystyle{unsrt}      
\bibliography{COXPH.bib}

\end{document}